\documentclass[12pt,preprint]{aastex}
\usepackage{amssymb}
\usepackage{stmaryrd}
\usepackage{amssymb,amsmath, epsfig, epsf, lscape}
\usepackage[dvips]{color}

\def\nar{New Astronomy Review}

\def\A{${\rm \AA}$}

\newcommand{\lbol}{\ensuremath{L\mathrm{_{bol}}}}

\newcommand{\ledd}{\ensuremath{L\mathrm{_{Edd}}}}
\newcommand{\lr}{\lbol/\ledd}

\newcommand{\kms}{\ensuremath{\mathrm{km~s^{-1}}}}
\newcommand{\mbh}{\ensuremath{M_\mathrm{BH}}}

\newcommand{\st}{\ensuremath{\sigma_{\rm \scriptscriptstyle T}}}

\newcommand{\hb}{H\ensuremath{\beta}}

\newcommand{\feii}{Fe\,{\footnotesize II}}

\newcommand{\mgii}{Mg\,{\footnotesize II}}
\newcommand{\heii}{He\,{\footnotesize II} }

\newcommand{\civ}{C\,{\footnotesize IV}}
\newcommand{\nv}{N\,{\footnotesize V}}

\newcommand{\siiv}{Si\,{\footnotesize IV}}
\newcommand{\ciii}{C\,{\footnotesize III]}}
\newcommand{\aliii}{Al\,{\footnotesize III}}
\newcommand{\siiii}{Si\,{\footnotesize III]}}
\newcommand{\feiii}{Fe\,{\footnotesize III}}

\newcommand{\colnh}{\ensuremath{N_\mathrm{H}}}

\def\gs{\mathrel{\raise1.16pt\hbox{$>$}\kern-7.0pt
\lower3.06pt\hbox{{$\scriptstyle \sim$}}}}
\def\ls{\mathrel{\raise1.16pt\hbox{$<$}\kern-7.0pt
\lower3.06pt\hbox{{$\scriptstyle \sim$}}}}

\slugcomment{To appear in {\it The Astrophysical Journal}}
\righthead{Two \civ~BELRs} \shorttitle{Two \civ~BELRs}
\shortauthors{Wang H.Y. et al.} 

\begin{document}
\title {Coexistence of Gravitationally Bound and Radiation Driven \\ \civ\ Emission Line Regions in Active Galactic Nuclei}
\author{Huiyuan Wang\altaffilmark{1,2},
Tinggui Wang\altaffilmark{1,2}, Hongyan Zhou\altaffilmark{1,2,6},
Bo Liu\altaffilmark{1,2}, Jianguo Wang\altaffilmark{3}, Weimin
Yuan\altaffilmark{4} and Xiaobo Dong\altaffilmark{1,2,5}}

\altaffiltext{1}{Key Laboratory for Research in Galaxies and
Cosmology, University of Science and Technology of China, Chinese
Academy of Sciences, Hefei, Anhui 230026, China;
whywang@mail.ustc.edu.cn} \altaffiltext{2}{Department of
Astronomy, University of Science and Technology of China, Hefei,
Anhui 230026, China} \altaffiltext{3}{National Astronomical
Observatories/Yunnan Observatory, Chinese Academy of Sciences,
P.O. Box 110, Kunming, Yunnan 650011, China}
\altaffiltext{4}{National Astronomical Observatories, Chinese
Academy of Sciences, Beijing 100012, China} \altaffiltext{5}{The
Observatories of the Carnegie Institution for Science, 813 Santa
Barbara Street, Pasadena, CA 91101, USA} \altaffiltext{6}{Polar
Research Institute of China, Jinqiao Rd. 451, Shanghai, 200136,
China}

\begin{abstract}
There are mutually contradictory views in the literature of the
kinematics and structure of high-ionization line (e.g. \civ)
emitting regions in active galactic nuclei (AGNs). Two kinds of
broad emission line region (BELR) models have been proposed,
outflow and gravitationally bound BELR, which are supported
respectively by blueshift of the \civ\ line and reverberation
mapping observations. To reconcile these two apparently different
models, we present a detailed comparison study between the \civ\
and \mgii\ lines using a sample of AGNs selected from the Sloan
Digital Sky Survey. We find that the kinematics of the \civ\
region is different from that of \mgii, which is thought to be
controlled by gravity. A strong correlation is found between the
blueshift and asymmetry of the \civ\ profile and the Eddington ratio. This
provides strong observational support for the postulation that the
outflow is driven by radiation pressure. In particular, we find
robust evidence that the \civ\ line region is largely dominated by
outflow at high Eddington ratios, while it is primarily
gravitationally bounded at low Eddington ratios. Our results
indicate that these two emitting regions coexist in most of AGNs.
The emission strength from these two gases varies smoothly with
Eddington ratio in opposite ways. This explanation naturally
reconciles the apparently contradictory views proposed in previous
studies. Finally, candidate models are discussed which can account
for both, the enhancement of outflow emission and suppression of
normal BEL, in AGN with high Eddington ratios.
\end{abstract}


\section{Introduction}\label{sec_intro}

It is now generally believed that super-massive black holes reside
in the center of active galactic nuclei (AGNs; Lynden-Bell 1969).
Through accretion of gas, AGNs release vast radiant energy,
including high-energy ionizing photons, from an accretion disk
around the black hole (Rees et al. 1984). The radiation
photoionizes and heats surrounding gas so that it is able to leave
its imprint, for example broad emission line (BEL) and broad
absorption line (BAL), on the emergent continuum spectra emitted
by accretion disk (Davidson \& Netzer 1979; Osterbrock 1989;
Weymann et al. 1991). The observed properties of these lines, such
as strength, width, and profile, are determined by the combined
effects of various physical processes. Hence, studies of these
line features can reveal the physical conditions and processes in
the central region of AGNs, and further shed light on the
accretion and radiation mechanisms, and the co-evolution of black
holes and their host galaxies.

The broad emission line region (BELR) is the most studied
structure of AGNs. Extensive efforts had been devoted to probing
its structure and kinematics, which are important for estimation
of black hole masses and understanding of line formation.
Velocity-resolved reverberation mapping of several local AGNs
shows that there is no significant difference in the temporal
response of the blue and red parts of the BELs to variations in the
continuum (Gaskell 1988; Koratkar \& Gaskell 1991; Korista et al.
1995; Done \& Krolik, 1996; O'brien et al. 1998). These studies demonstrate
that the predominant motion of the BELR is either Keplerian or
virial motion, both driven by the gravity of the central black
hole (see Gaskell 2009, for a review). Further robust evidence
supporting this view is given by the correlation between the BELR
size and line width in the form $r \propto \sigma^{-2}$ for
various emission lines found in a few well-studied AGNs (Krolik et
al 1991; Peterson \& Wandel 1999; 2000; Onken \& Peterson 2002).
This correlation is readily expected for gravitation dominated
kinematics. Particularly, these works showed that the emitting
regions of both high- and low-ionization lines (e.g. \civ\ and
\hb) are gravitationally bound.

However, studies of luminous AGNs at high redshifts uncover quite
different behaviors of the BELR, especially the high-ionization
line (e.g. \civ) region. Gaskell (1982; see also Wilkes 1984;
Marziani et al. 1996) found that the peak of the \civ\ line tends
to be blueshifted with respect to the peak of low-ionization lines
in a small sample of high-redshift AGNs. This was confirmed by
Richards et al. (2002; also Vanden Berk et al. 2001) using a large
sample of AGNs selected from the Sloan Digital Sky Survey (SDSS;
York et al. 2000). Subsequently, the \civ\ blueshift was detected
in several low-luminosity AGNs (e.g. Leighly \& Moore 2004).
The blueshift is difficult to
reconcile with gravitationally bound BELR models, but is
considered as a signature of outflowing gas (Gaskell 1982;
Marziani et al. 1996; Leighly 2004). Outflow models based on
dynamical and photoionization calculations have been put forward
to reproduce the observed line profiles, equivalent widths and
line ratios, and further imposing constraints on the density,
ionization state and geometry of the line emitting gas (e.g.
Murray \& Chiang 1997; Leighly 2004; Wang et al. 2009b). In fact,
the frequent occurrences of blueshifted narrow absorption lines
and BALs in AGN spectra have proved that outflow is ubiquitous in
AGN (Weymann et al .1991; Crenshaw, Kraemer \& George 2003).
Outflow is therefore a natural interpretation of blueshift of
high-ionization lines.

Although there have been a few attempts to tune one single model,
i.e. outflow or gravitationally bound BELR, to simultaneously
account for the contradictory observational results in
reverberation-mapped objects and blueshifted \civ\ AGNs (e.g.
Chiang \& Murray 1996; Gaskell \& Goosmann 2008), the dramatic
inconsistency may indeed manifest fundamental differences in the
structure and kinematics of these two line emitting regions. If
this is the case, it would be interesting to find out what causes
such differences. Previous studies have already provided some
meaningful observational constraints on the underlying processes.
For example, it is well known that the blueshift of the \civ\
emission line in radio quiet AGNs is, on average, stronger than in
radio loud AGNs (Marziani et al. 1996; Sulentic et al. 2000a;
Richards et al. 2002; 2011). Recently, several studies found that
the \civ\ blueshift decreases with the increase of the X-ray to UV
flux ratio (Gibson, Brandt \& Schneider 2008; Richards et al.
2011). In addition, Sulentic et al. (2007) found that the
correlation between the width and equivalent width of the \civ\
BEL varies dramatically with the width of the \hb\ BEL. Richards
et al. (2011) found that \civ\ blueshift varies with the ionizing
spectral energy distribution. All these facts point towards that
BELR structure is related to the accretion process of the central
engine.

It is worthwhile to note that the low-ionization lines, such as
\mgii\ and \hb, are fairly close to the systemic redshift
(Marziani et al. 1996; Sulentic et al. 2000a; Richards et al.
2002). Apparently different from the high-ionization lines, the
low-ionization line region is gravitationally bound, and outflow
model is unlikely. This is supported by the fact that local AGNs
follow the same luminosity-BELR size relationship for the \hb\
line as for those well-studied AGNs, whose BELR have been
demonstrated to be governed by gravity (Kaspi et al. 2005; Bentz
et al. 2006). In particular, the power-law slope of this
relationship is about 0.5, indicating that all of these AGNs have
roughly the same ionization state and gas density in the
low-ionization lines region. The \hb\ and \mgii\ lines are
therefore widely adopted to estimate the black hole masses, \mbh,
of AGNs (Kaspi et al. 2000; Collin et al. 2006; Vestergaard \&
Peterson 2006; Mclure \& Jarvis 2002; Onken \& Kollmeier 2008;
Wang et al. 2009a). The \mbh\ estimate is plagued by many
uncertainties in, for example, the geometry, kinematics and
inclination of the BELR, the measurement of line with, the
influence from other forces (such as radiation pressure), the
validity of extrapolating the small-sample results to a large
sample (Krolik 2001; Collin et al. 2006; Marconi et al. 2008;
Richards et al. 2011). Although the \mbh\ estimate can be
calibrated using the correlation between \mbh\ and bulge/spheroid
stellar velocity dispersion (e.g. Onken et al. 2004), it is
accurate only from a statistical point of view.

In this paper, we probe the structure and kinematics of the
high-ionization line emitting region via a comparison with the
low-ionization lines. We use the \civ\ line, the most prominent
metal line, to represent high-ionization lines, while we adopt \mgii\
as a representation of low-ionization lines. Both lines are
observable in SDSS AGNs at redshift $z\sim2$, from which a
sufficiently large sample can be obtained for our purpose. Please
see Section \ref{sec_sam} for the sample selection. We present in
Section \ref{sec_c4b} the correlations between the properties of
\civ\ line, and the composite spectra. In Section \ref{sec_mc}, we
examine whether the \civ\ and \mgii\ line properties follow the same
correlations. Similarity in the correlations indicates that two
lines come from regions with similar kinematics, while opposite
results mean that the structures are different. Our method differs
from earlier investigations, in which the authors directly
compared the properties of the two lines. Then, in Section
\ref{sec_dis}, we analyze the geometry and kinematics of the \civ\
region and discuss the candidate models. We also present a
comprehensive comparison between BAL and the blueshifted \civ\ BEL.
Finally, we summarize our results in Section \ref{sec_sum}.

\section{Sample and Data Analysis}\label{sec_sam}

We select AGNs in the redshift range $1.7<z<2.2$ from the Fifth
Data Release (DR5) of SDSS spectroscopic database (Schneider et
al. 2007). The redshift range is chosen in such a way that both the \civ\ and
\mgii\ lines fall in the wavelength coverage of the SDSS
spectrograph. To ensure reliable measurements of emission line
parameters, we only select objects with median signal-to-noise
ratio (S/N)$\geq$ 7 per pixel in both the \civ\ (1450-1700 \A) and
the \mgii\ (2700-2900 \A) spectral regions. We further discard the
broad absorption line AGNs as cataloged by Scaringi et al. (2009).
6009 AGNs meet these criteria. The SDSS spectra are corrected for
the Galactic extinction using the extinction map of Schlegel et
al. (1998) and the reddening curve of Fitzpatrick (1999), and
transformed to the rest frame using the improved redshifts for
SDSS AGNs as computed by Hewett \& Wild (2010, hereafter HW10). To measure the
broad lines, we perform continuum and emission-line fitting using
Interactive Data Language (IDL) code based on MPFIT (Markwardt
2009), which performs $\chi^2$-minimization by the
Levenberg-Marquardt technique.

The \mgii\ broad lines are fitted using the exactly same method
as Wang et al. (2009a). Here we only present a brief
description of the procedures. First, we simultaneously fit the
featureless continuum (assumed to be a power law) and the \feii\
multiplet emission, which together constitute the so-called
pseudocontinuum. \feii\ is modelled with the tabulated
semi-empirical template generated by Tsuzuki et al. (2006) based
on their measurements of I ZW1; in the wavelength region covered
by the \mgii\ emission, this template uses the calculation with
the CLOUDY photoionization code (Ferland et al. 1998). After
subtracting the fitted pseudocontinuum, the broad components of
the \mgii\ $\lambda\lambda$2796, 2803 doublet lines are each
modelled with a truncated five-parameter ($p_i$, i=0 to 4) Gauss-Hermite series
profile (van der Marel \& Franx 1993). The expression for a Gauss-Hermite function is
\begin{eqnarray}
&{\rm G_h}(x)=p_0e^{-x^2/2}[1+p_3h_3(x)+p_4h_4(x)],\nonumber\\
&h_3(x)=\frac{1}{\sqrt{6}}(2\sqrt{2}x^3-3\sqrt{2}x),\nonumber\\
&h_4(x)=\frac{1}{\sqrt{24}}(4x^4-12x^2+3).\nonumber
\end{eqnarray}
where $x=(\lambda-p_1)/p_2$. The narrow component of each line is fitted with a single Gaussian. The full width at half
maximum (FWHM) is measured from the Gauss-Hermite model of \mgii\
$\lambda$2796, and rest equivalent width (EW) is the sum of the
broad doublet lines.

To measure the \civ\ broad lines, we first fit the local
continuum with a power-law in two wavelength windows near 1450\A\
and 1700\A, that have little or no contamination from
emission-lines. After subtracting the continuum, we fit the
residual spectrum around \civ\ with two Gaussians. Since the red
wing of \civ\ is contaminated by \heii, only the spectral region
of 1450-1580\A\ is considered in fitting the \civ\ line. We also
attempted to fit the residual spectrum with three Gaussians, which
results in little or no significant improvements.
We thus adopt the two-Gaussian model here.

In this work we introduce a blueshift and asymmetry index (BAI) to
measure the deviation of the \civ\ line from an unshifted and
symmetric profile. BAI is defined as the flux ratio of the blue
part to the total profile, where the blue part is the part of
\civ\ line at wavelengths short of 1549.06\A, the laboratory
rest-frame wavelength of \civ\ doublets. The line parameters, BAI,
FWHM and EW, of \civ\ are measured from the composite profiles of
the two-Gaussian model (in the wavelength range of 1450-1700\A).
The BAI estimation may be affected by the accuracy of the
redshifts which we use to transform the observed spectra to the
rest frame. To reduce possible uncertainty of BAI thus introduced,
we adopt the redshifts provided by HW10\footnote{More recently,
Richards et al. (2011) also used these redshifts to compute the
\civ\ line shift. They found that the resultant line-shift is
consistent with the shift with respect to the low-ionization line
\mgii, which is often used in the literature.}, which were derived
by cross correlation of observed spectra with a
carefully-constructed template. In particular, they corrected for
the dependence of emission-line shift on luminosity and redshift,
which is relevant to our investigation. BAI measures the combined
effects of the asymmetry and shift of a line profile, which are
generally treated separately in the literature (e.g. Sulentic et
al. 2000a). In the Appendix we consider line shift and asymmetry
separately and conclude that using BAI is a robust choice for our
purposes. We use blueshift and BAI alternately in this paper.

The fitting results for most of the objects are reasonable
according to our subsequent visual inspection. However, a small
fraction of objects display strong narrow absorption in the
emission lines regions, especially \civ\ spectral region, and cannot
be well fitted by our automated processing. To minimize
outliers left by the procedures, we eliminate objects with
unacceptable fitting, i.e. $\chi^2/$d.o.f$>1.5$. This restriction
leads to a final working sample of 4963 AGNs. We note that the last selection
has little or no effect on the results obtained in this paper.

\section{Blueshift of \civ\ Emission Line}\label{sec_c4b}

We show the distribution of BAI for the entire sample in Fig.
\ref{fig_udis}. This parameter distributes in a wide range, from
about 0.4 to more than 0.9, with a median value of 0.63. According
to our definition of BAI, a blueshifted line has a BAI parameter
larger than 0.5. Obviously, most of the AGNs in our sample display
blueshifted \civ\ emission lines with respect to their
systemic redshifts, in agreement with previous findings (Gaskell
1982; Richards et al. 2002). Only 7.2\% of the AGNs are
redshifted outliers. This fraction is the same as that found in
Richards et al. (2011), in spite of that we adopt a different method to
denote the blueshift. Since the blueshifted \civ\ line has been
suggested to be emitted from outflows (e.g. Gaskell 1982; Murray
\& Chiang 1997; Leighly 2004), the BAI distribution suggests that
outflows are common in AGNs.

We then show the equivalent width of \civ\ as a function of BAI
and FWHM(\civ) in the two upper panels of Fig. \ref{fig_uwe}. The
\civ\ emission line is weaker for AGNs with larger blueshifts or
line widths. Both two trends are consistent with previous results.
Richards et al. (2011) found an anti-correlation between blueshift
and equivalent width (see also Marziani et al. 1996; Leighly \&
Moore 2004), and Wills et al. (1993) reported an anti-correlation
of FWHM with EW. The left panel of Fig. \ref{fig_uf} shows
FWHM(\civ) against BAI. Our result confirms the finding of
Richards (2006) that the line width of \civ\ is, on average,
slightly larger for AGNs with larger blueshifts. Inspecting the
scatter plot in detail, one can find that the trend is complex and
there appears to exist a `V'-shaped lower envelope. It implies a
mixture of origins of the \civ\ emission. We will come back to
this issue later.

Composite spectra can provide a wealth of diagnostic information.
We show the composite spectra as a function of BAI in Fig.
\ref{fig_cs1}. Our composites are constructed through the
arithmetic mean method. We first sort the AGN sample according to
their BAI then sub-divide it into four equally-sized subsamples
(each subsample contains about 1241 AGNs). For each AGN, we use
the redshift from HW10 to deredshift the
spectrum. The spectrum is then normalized at 1450\A, rebinned into
the same wavelength grids. The composite spectra are created by
combining these normalized spectra in each of these subsamples.
Only four major emission line regions, \siiv, \civ, \ciii\ and
\mgii, are presented. The spectra are normalized in the
local continuum regions: (1355,1450),
(1450,1700), (1830,1975), (2695,2955) (in units of \A) for \siiv,
\civ, \ciii\ and \mgii.

As expected, the \civ\ emission line in our composites shifts
significantly towards shorter wavelength as BAI increases. The
equivalent width of \civ\ decreases with increasing BAI,
consistent with Fig. \ref{fig_uwe}. The blueshift is also
remarkable for \heii\ and moderate for \siiv, but seems to be
absent in lower-ionization lines, such as \mgii\ and \aliii, in
good agreement with Richards et al. (2002). We also create
composites as a function of EW(\civ) and BAI, following Richards
et al. (2011). The composites are very similar to that shown in
Richards et al. This is not surprising since we use the same AGN
redshifts provided by HW10. Our composites reveal a significant
excess flux in the blue wing of \civ\ (around 1500 \A), the
strength of which increases with BAI (Fig. \ref{fig_uwe}). We then
take the difference between the largest-BAI and smallest-BAI
composite spectra and show the difference as a function of
velocity with respect to \civ\ in Fig. \ref{fig_ex} (thick line).
The blue excess wing peaks at 8000\kms, ranging between
$\sim4000$\kms and 11000\kms.

Two additional phenomena are worth noting. First, the mean width
of \mgii\ tends to be narrower for AGNs with larger \civ\
blueshifts. We will address this issue in next section. Second,
the equivalent width of \aliii\ is dramatically boosted at large
BAI, different from other lines. Richards et al. (2011) paid
particular attention to the unusual property of the \aliii\ line.
They suggested that the strong \aliii, together with the large
\siiii/\ciii\ ratio, is indicative of an X-ray weak spectrum for
these AGNs. Although this explanation is reasonable, it might
simply be that \aliii\ is contaminated by other lines. One
possible contamination comes from the UV \feiii\ complexes. The
\feiii\ complexes are usually more prominent on the
long-wavelength side of \ciii (Laor et al. 1997). However, we do
not find the expected excess at the corresponding wavelength and
thus rule out this probability. Another promising explanation is
that this excess flux is the \ciii\ emission of the outflow. To
demonstrate this, we show the difference spectra of the
largest-BAI and smallest-BAI composites as a function of velocity
with respect to \ciii\ in Fig. \ref{fig_ex}(thin line). One can
find the average velocity of the `excess of \aliii' is about
8000\kms. This velocity is very similar to that of the blue wing
excess of \civ\, and thus supports our interpretation. Difference
spectra are usually sensitive to the normalization. To investigate
into this, we also try to use other local continuum regions. The
velocity ranges of the excess flux for both \civ\ and \ciii\
change only slightly.

The blueshifted \ciii\ component suggests that the number density
of the large-velocity part of the \civ\ BELR cannot be much larger
than $3\times10^{9}{\rm cm}^{-3}$, which is the critical density
of the intercombination line \ciii\ (Osterbrock 1989). This value
is consistent with that estimated by Wang et al.(2009b), but much
lower than that estimated by Ferland et al. (1992), about
$10^{11}{\rm cm}^{-3}$ (see also Kuraszkiewicz et al. 2000). The
discrepancy can be understood if there exist two kinds of \civ\
emitting region: a normal gravitationally bound BELR and an
outflow, as we will discuss in this paper. In fact, studies of
blueshifted absorption lines revealed quite low densities in
outflows (see Crenshaw et al. 2003 for a review). Moreover, the
column density of outflow can be constrained based on the absence
of blueshifted component in low-ionization line. For example,
Leighly (2004) performed photoionization calculation to model the
outflow components of various UV lines in two narrow-line Seyfert
1 galaxies and derived a rather low column density: $\log \colnh
\simeq 21.4$ for outflow gas. Leighly (2004) did not consider the
blueshifted \ciii\ component. If taking into account this
component, one may get a slightly larger column density. The lower
density and column density than the typical values of normal
emitting gas may be due to the expansion of outflow.

\section{Comparison of Kinematics of \mgii\ and \civ}\label{sec_mc}

In this section, we examine whether the \mgii\ and \civ\ line
properties follow the same correlations. Since \mgii\ is thought
to be emitted from a gravitationally bound BELR, the comparison
may offer insight into the kinematics and geometry of the \civ\
emitting region. Note that our method differs from earlier
investigations, in which the authors directly compared the
properties of the two lines.

\subsection{Correlations for \mgii\ and Fundamental Driver of \civ\ Blueshift}

We show the rest equivalent width of \mgii\ against BAI and
FWHM(\mgii) in the lower panels of Fig. \ref{fig_uwe}. Note that
the BAI presented here is calculated using \civ\ line. One can see
that EW(\mgii) declines with increasing BAI. The trend is similar
to but slightly weaker than \civ. Of particular interest is the
positive correlation between EW(\mgii) and FWHM(\mgii), which is
in contrast to \civ (Fig. \ref{fig_uwe}). We also found such a
correlation in a low-redshift AGN sample (Dong et al. 2009). The
opposite behaviors indicates that the properties of \mgii\ and
\civ\ emitting regions are essentially different.

The scatter plot of FWHM(\mgii) versus \civ\ blueshift is shown in
the right panel of Fig. \ref{fig_uf}. Here again, we find a
different correlation from \civ. AGNs with narrow \mgii\ BEL have
a strong tendency to exhibit \civ\ emission with prominent
blueshift. This trend is appreciable in previous work based on
small low-redshift samples. For instance, Bachev et al. (2004)
found the \civ\ blueshift can only be found in objects with
FWHM(\hb) less than 4000\kms. Baskin \& Laor (2005) found an
marginal correlation between \civ\ blueshift and FWHM(\hb) in 81
PG QSOs. Sulentic et al. (2007) used the centroid at half maximum
of \civ\ profile as a surrogate of line shift. Their Figure 2
clearly shows a somewhat weak correlation between the blueshift
and FWHM(\hb) (see also Sulentic et al. 2000b). Since FWHM(\mgii)
is closely correlated with FWHM(\hb)(e.g. Wang et al. 2009a),
their results broadly agree with ours.

This correlation hints that the \civ\ blueshift is
driven by some primary physical parameters, such as
black hole mass and the Eddington ratio\footnote{\lbol\ is the
bolometric luminosity and \ledd\ is the luminosity required for
radiation pressure arising from electron scattering to balance the
gravity of the central black hole.} (\lr). We calculate black hole
masses based on the FWHM(\mgii) and the monochromatic luminosity
$L_{\rm 3000}=\lambda L(3000 {\rm \AA})$ using the formula of Wang
et al. (2009a)\footnote{Our tests show that the results presented
below do not change if we adopt other \mgii\ black hole mass
formulae.}. We then compute the Eddington ratio assuming a
constant bolometric correction, $L_{\rm bol}\simeq 5.9 L_{\rm
3000}$ (McLure \& Dunlop 2004). To give a quantitative analysis,
we perform Spearman rank correlation tests over the correlations
of BAI with \lr, \mbh\ and FWHM(\mgii), but note that these
properties are degenerate, with \lr$\propto$FWHM$^{-2}$, and
\mbh$\propto$FWHM$^{2}$. The correlations with \lr\ and
FWHM(\mgii) are comparable in strength, with Spearman correlation
coefficients $r_s$ of 0.54 and -0.50 respectively (the
probabilities for null hypothesis are less than $10^{-100}$), and
are both significantly stronger than the dependence on \mbh ($r_s$
= -0.39). Eddington ratio is therefore a more fundamental driver
of the \civ\ blueshift than black hole mass. In Fig. \ref{fig_uledd},
we show BAI as a function of \lr. As \lr\
increases from 0.1 to 1, the median value of BAI significantly
increases from $\sim$0.5 to $\sim$0.75. A commonly accepted view
for this correlation is that outflow is driven by radiation
pressure (e.g. Murray et al. 1995; Boroson 2002), and therefore
prominent for high \lr\ AGNs. If outflow carries away a significant amount of
the angular momentum (e.g. Wang et al. 2007), it can serve as a driver
of the accretion process, and consequently strengthen this
correlation.

The correlation of BAI with FWHM(\mgii) is as strong as that with
the Eddington ratio. It would be interesting to know which
physical property dominates the correlations involving BAI.
However, given the tight correlation between FWHM(\mgii) and the
Eddington ratio ($r_s$=-0.93) in our sample, that is flux limited,
it is impossible to disentangle these dependencies. A sample of
AGNs covering a wide range in luminosity is required for this
purpose. Recently, Richards et al. (2011) reported that the
reverberation mapped AGNs occupy only part of the \civ\ line
parameter space, and suggested that these objects have different
ionizing spectral energy distribution (SED) from the mean SED of
AGNs. It is unclear whether applying the scaling relation to the
whole sample would lead to systematic error in determining \mbh\
and \lr. If yes, the origin of the correlation between BAI and the
inferred \lr\ would be more complex than that we discussed above,
because the ionizing SED also correlates with the \civ\ blueshift
(Richards et al. 2011). Even so, it doesn't necessarily mean the
relationship between real \lr\ and BAI is weaker than that shown
in Fig. \ref{fig_uledd}. In fact, the correlation between SED,
characterized by the ratio of X-ray flux to UV flux
($\alpha_{ox}$), and the inferred \lr\ and line width is weak or
absent (Vasudevan \& Fabian 2007; Shemmer et al. 2008). These
results doesn't favor that the estimates of \lr\ and \mbh\ are
significantly affected by the assumption of a single mean SED.
However, $\alpha_{ox}$ cannot fully characterize the ionizing SED,
further work is needed. Moreover, the constant bolometric
correction may introduce systemic bias in the dependence of BAI on
\lr. But it is not important here because the bolometric
corrections differ between objects by a factor of 2 (e.g. Richards
et al. 2006), much less than the \lr\ range of our sample.

\subsection{\civ\ in High and Low \lr\ AGNs}

When considering the correlations involving FWHM, we can see very
different or even opposite behaviors between \civ\ and \mgii. FWHM
may be a key to understanding the difference between \civ\ and
\mgii\ emitting regions. The width of an emission line reflects
the motion of the corresponding emitting gas. \mgii\ gas is
thought to be gravitationally bound, whereas \civ\ line is
generally blueshifted and the corresponding emitting gas may be
impacted or driven by the radiation field. Thus, FWHM(\mgii) is a
measure of the gravity, while FWHM(\civ) is more likely the result
of the competition between gravity and radiation pressure. It is
therefore not surprising to find inconsistent trends. In order to
understand the detailed effect of radiation pressure on \civ\ gas,
it is necessary to explore more differences between \mgii\ and
\civ, and to examine whether these trends for \civ\ hold for AGNs
of different Eddington ratio.

To do so, we select two extreme subsamples which comprise the 25\%
highest \lr\ and 25\% lowest \lr\ AGNs, designated as sample A and
B respectively. We first show the FWHM(\civ) as a function of BAI
for these two subsamples in the left panels of Fig. \ref{fig_fb}.
There is a clear trend that \civ\ line width increases with BAI in
sample A, but this trend disappears in sample B. An evidence for
this discrepancy can also be found in the entire sample (left
panel of Fig. \ref{fig_uf}). For comparison, we also show
FWHM(\mgii) versus BAI for the \emph{entire} sample in the same
two panels (black points) as background. The \civ\ line of AGNs in
sample A reveals an opposite tendency compared to the
low-ionization line \mgii, while the \civ\ distribution of sample
B is well consistent with the upper half of the \mgii\
distribution. We then make the same analysis of the FWHM-EW
correlations for both lines and obtain similar results (see the
right panels of Fig. \ref{fig_fb}). On one hand, AGNs in sample A
exhibit a strong anti-correlation between FWHM(\civ) and EW(\civ),
which is completely opposite to the correlation for \mgii. This
correlation also exists in the whole sample, albeit weaker. On the
other hand, for sample B AGNs, the correlation becomes marginal
and the discrepancy between \civ\ and \mgii\ become tiny compared
to sample A.

We note that the results shown above do not change if we select
subsamples based on FWHM(\mgii) rather than \lr. That is again due to the fact that
FWHM(\mgii) is strongly correlated with Eddington ratio in our
flux limited sample. Using 130 low-redshift AGNs with HST
ultraviolet-band spectroscopic observation, Sulentic et al.(2007)
found FWHM(\civ) increases with the blueshift \emph{only} in AGNs
with FWHM(\hb) $\leq$4000\kms. Their results are consistent with
ours based on a much larger sample and \mgii, supporting that the
kinematics of \hb\ and \mgii\ regions are similar. They found AGNs
with FWHM(\hb) less and greater than 4000\kms\ exhibit different
FeII and radio properties and further suggested that there is
apparent dichotomy between these two populations(see also Sulentic
et al. 2000a;2000b; Bachev et al. 2004; Zamfir, Sulentic \&
Marziani 2008). We also find similar differences between sample A
and B AGNs in the parameter space defined by BAI, FWHM and EW (and
the same differences between the small and large FWHM(\mgii)
subsamples). \emph{More importantly, we find  that the \civ\
emission of sample A is totally different from  the low-ionization
line \mgii, while  the \civ\ emission of sample B is similar to
\mgii}. In addition, Fig. \ref{fig_uf} and \ref{fig_uledd} show
clearly that the \civ\ blueshift varies continuously and smoothly
with FWHM(\mgii) and \lr. This implies that the corresponding
variation in the properties of the \civ\ emitting region is
smooth, rather than abrupt.

\section{Discussion}\label{sec_dis}

In this section, we analyze the kinematics and geometry of the
\civ\ region on the basis of the results presented above. Then we
discuss candidate BELR models that can account for the specific
behavior of \civ. To get more insight into the origin of the \civ\
blueshift, we finally present a comprehensive comparison between
the \civ\ blueshift and BAL.

\subsection{Gravitationally Bound and Radiation Driven Emission
Line Regions}

The \civ\ emission of the AGNs in sample\,A is strongly
blueshifted, and it also displays an opposite tendency compared to
\mgii\ in the parameter space defined by BAI, FWHM and EW. It is
very likely dominated by outflow, whose velocity range can be
roughly measured by FWHM(\civ). Further evidence for the outflow
origin of the \civ\ blueshift is given by comparison of the
blueshifted  \civ\ and BAL (see Sect. \ref{sec_cbo}), that is
believed to be produced by outflows.

The correlation between FWHM(\civ) and BAI found from sample A can
be easily understood in terms of outflows. In general, an outflow
with a large terminal velocity tends to produce emission with a
large blueshift and FWHM(\civ). However, not all outflow models
can successfully reproduce such a relationship. One of the
commonly proposed models is disk-like equatorial outflows with a
small opening angle. In this model, outflows on the far and near
sides emit redshifted and blueshifted \civ\ photons, respectively,
at comparable amounts (see, e.g. Figure 7 of Murray \& Chiang
1997). Only when viewed along the pole, the emission line is
significantly blueshifted. Meanwhile, we will see a very narrow
line because the line of sight is nearly perpendicular to the
outflow velocity. It is inconsistent with our finding here. One
solution is that, in the case of disk-like outflows, a large
opening angle is required. Alternatively, the outflow is funnel
like (see the geometry shown in Elvis 2000 and Wang et al. 2007),
even close to the polar direction in some cases (e.g. Zhou et al.
2006; Wang et al. 2008). For a funnel-like outflow, the projected
velocity of the far-side along the line of sight is so small that
it contributes only little to the red side of the \civ\
emission\footnote{The emission from an counter outflow on the
other side of a presumed optically thick accretion disk is
blocked.}.

The strong anti-correlation between FWHM(\civ) and EW(\civ) in
sample A suggests that the \civ\ emission is suppressed in
outflows at high velocities. This can possibly be ascribed to
that, with expansion, the density of high velocity outflow gets
lower, in agreement with our finding that the blueshifted
component of the \ciii\ emission is possibly enhanced in AGNs with
large BAI (see section \ref{sec_c4b} and Fig. \ref{fig_cs1}).
However, as a major coolant of BELR, \civ\ may be insensitive to
the change of the density. Alternatively, Murray \& Chiang (1997)
suggested that this correlation is related to the distribution of
the launching radius of outflows. Detailed dynamical and
photoionization model is needed to understand the underlying
process.

Different from the case of sample A, the \civ\ emission of the
AGNs in sample B nearly overlaps with \mgii\ in the parameter
space. In particular, the two lines have similar widths. To
demonstrate this more clearly, we show the probability
distribution of $\log({\rm FWHM(\mgii)/FWHM(\civ)})$ in the left
panel of Fig. \ref{fig_fr}, which peaks at zero, albeit a large
scatter of 0.16 dex. For comparison, we also plot the result for
sample A,  show that \civ\ is much broader than \mgii. The
similarity between the widths of \civ\ and \mgii\ suggests that
they largely come from the same emitting gas, which should be
optically thick to ionizing radiation. Considering that radiation
pressure is not important for optically-thick clouds (e.g. Fabian
et al. 2006; Marconi et al. 2008) and that \mgii\ is a reasonable
indicator of black hole mass, our results suggest that the
kinematics of the \civ\ emission-line gas in these AGNs is
primarily governed by gravity rather than by radiation field. This
argument is also consistent with the facts that sample B AGNs have
small blueshifts of \civ\ and weak radiation field compared to
gravity for their low \lr.

The most convincing evidence to date for the gravitationally
dominated kinematics of the \civ\ regions, from reverberation
mapping observations of  local AGNs, includes similar response
timescales of the blue and red wings of \civ, the correlation
between the BELR size and line width (Gaskell 1988; Koratkar \&
Gaskell 1991; Korista et al. 1995; Peterson \& Wandel 1999; 2000;
Onken \& Peterson 2002). We examine the Eddington ratio of these
reverberation mapped objects listed in Koratkar \& Gaskell (1991),
Peterson \& Wandel (2000), and Onken \& Peterson (2002) by
collecting data from the literature, and find that almost all of
these have quite low Eddington ratio, generally less than 0.1 (see
Wang et al. 2009a for black hole mass and luminosity data).
Moreover, radio-loud AGNs have averagely lower Eddington ratios
(e.g. Boroson 2002; Zamfir et al. 2008) and weaker \civ\
blueshifts in comparison with radio-quiet AGNs. Both results are
in agreement with our conclusion that the \civ\ region tends to be
gravitationally bound at low \lr.

\begin{figure}
\epsscale{1}\plotone{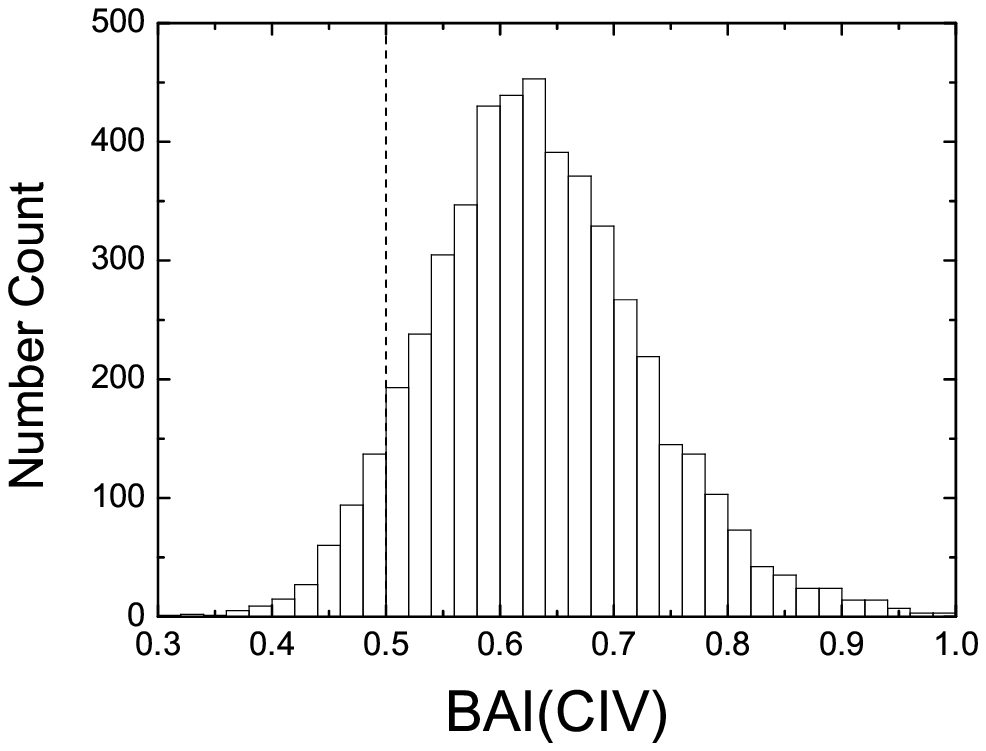} \caption{The
distribution of blueshift and asymmetry index (BAI) for \civ~BELs. The
dashed line indicates an un-blueshifted \civ~line
profile.}\label{fig_udis}
\end{figure}

\begin{figure}
\epsscale{1}\plotone{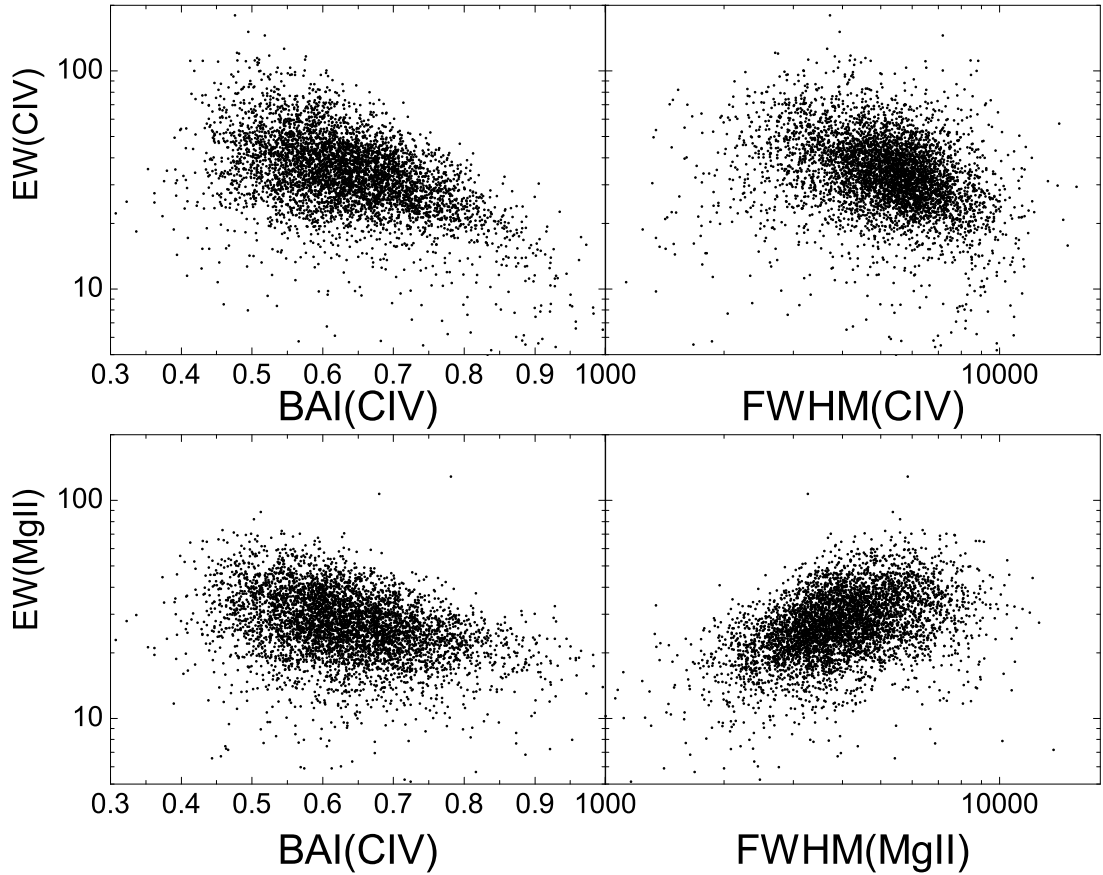} \caption{The rest equivalent
width as a function of the full width at half maximum (FWHM) and
BAI(\civ). The upper panels show the results for \civ, while the
lower panels are for \mgii. Note that, in the lower left panel,
BAI is calculated using the \civ\ BEL.}\label{fig_uwe}
\end{figure}

\begin{figure}
\epsscale{1}\plotone{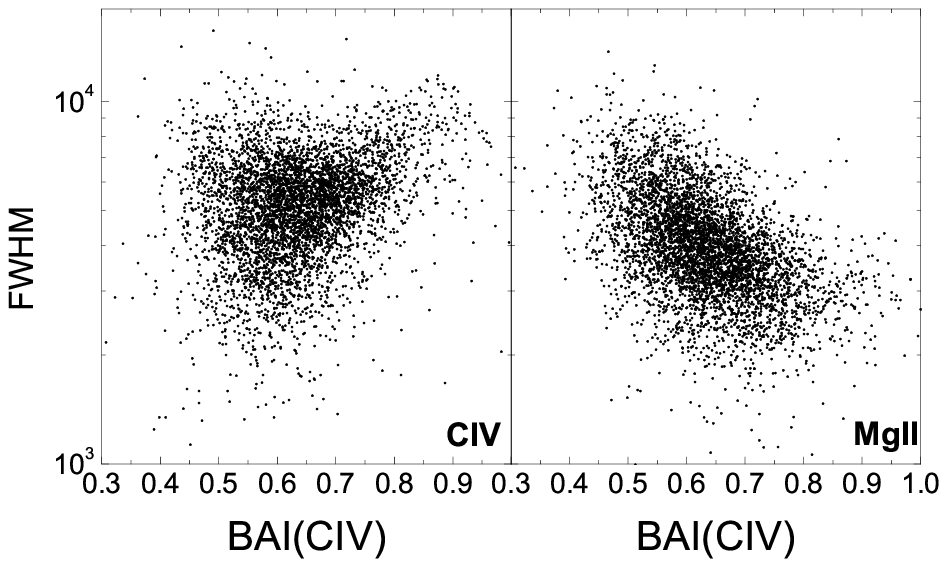} \caption{The left panel shows
FWHM(\civ) versus BAI(\civ), while the right panel show
FWHM(\mgii) versus BAI(\civ).}\label{fig_uf}
\end{figure}

\begin{figure}
\epsscale{1}\plotone{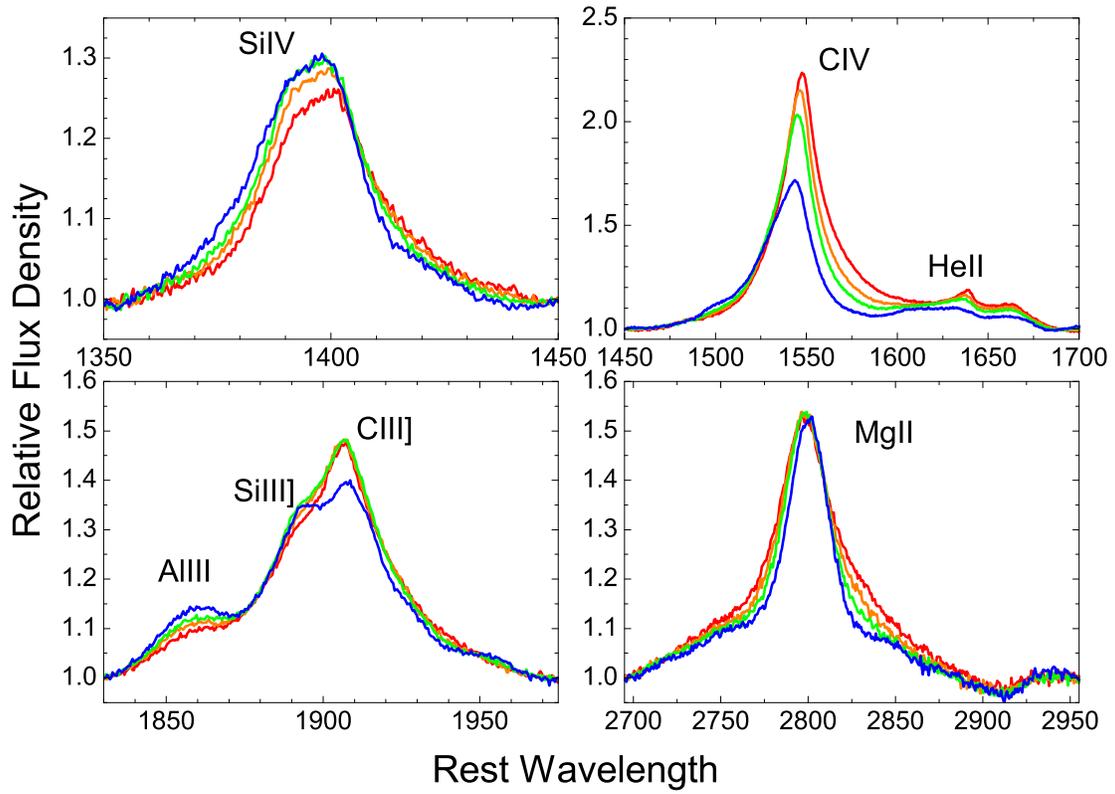} \caption{Strong emission lines
region of the composite spectra. The spectra, in the order of
increasing BAI(\civ), are plotted as red, orange, green and blue,
respectively.}\label{fig_cs1}
\end{figure}

\begin{figure}
\epsscale{1}\plotone{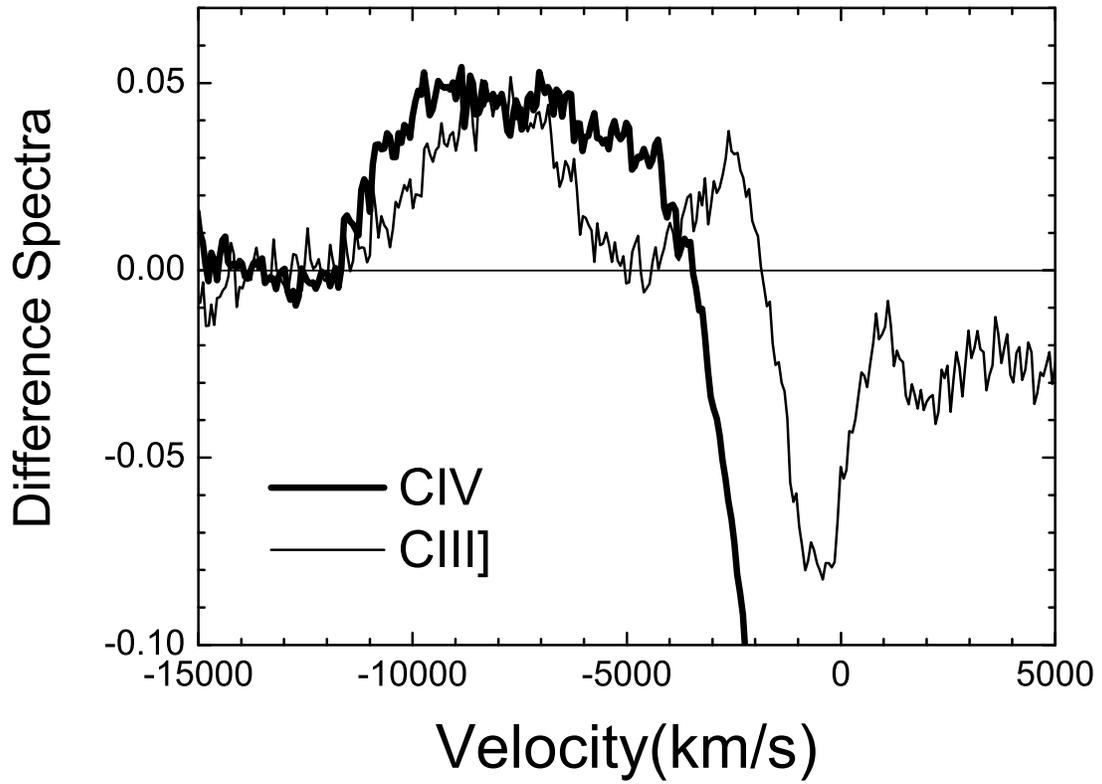} \caption{The difference
between the largest-BAI(\civ) and smallest-BAI(\civ) composite
spectra as a function of velocity with respect to both \ciii(thin
line) and \civ(thick line) respectively. Please see text for
details.}\label{fig_ex}
\end{figure}

\begin{figure}
\epsscale{1}\plotone{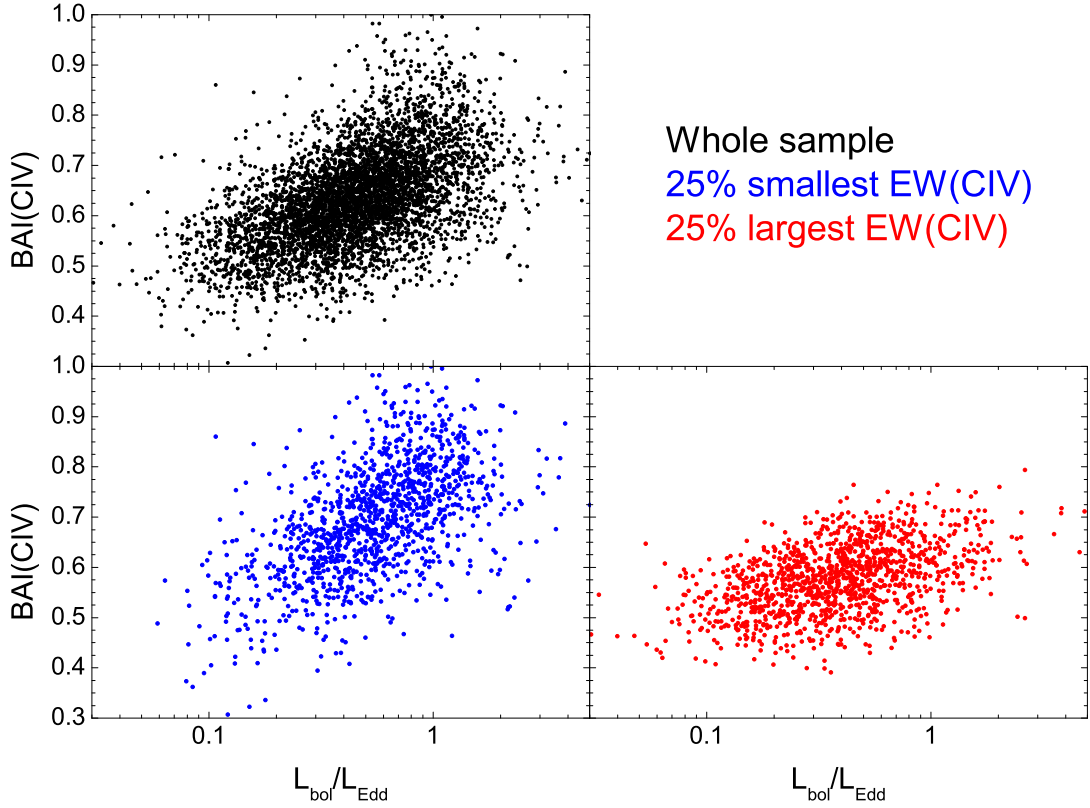} \caption{The \civ\ BAI as a
function of the Eddington ratio, \lr. The black hole mass and
Eddington ratio is calculated based on \mgii\ mass formula in Wang
et al. (2009a). The black points show the results for the whole sample.
The red points are for 25\% largest EW(\civ) AGNs, while
the blue points are for 25\% smallest EW(\civ) AGNs. For clarity, we don't combine these panels.}\label{fig_uledd}
\end{figure}

\begin{figure}
\epsscale{1}\plotone{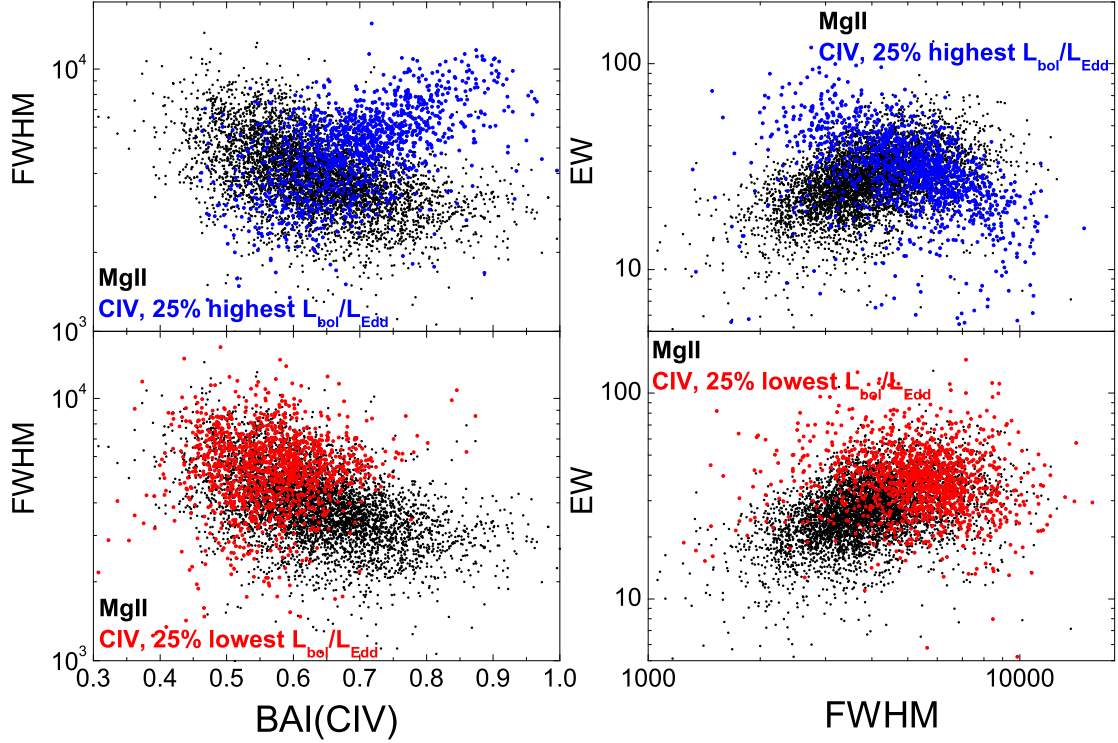} \caption{Left panels: FWHM
versus BAI(\civ). Right panels: EW versus FWHM. The black points
represent \mgii\ line, while the colored points are for \civ. The
red points are for 25\% lowest \lr\ AGNs (sample B), while the
blue points are for 25\% highest \lr\ AGNs (sample A). The black
points are for the whole sample. For clarity, we don't combine
these panels.}\label{fig_fb}
\end{figure}

\begin{figure}
\epsscale{1}\plotone{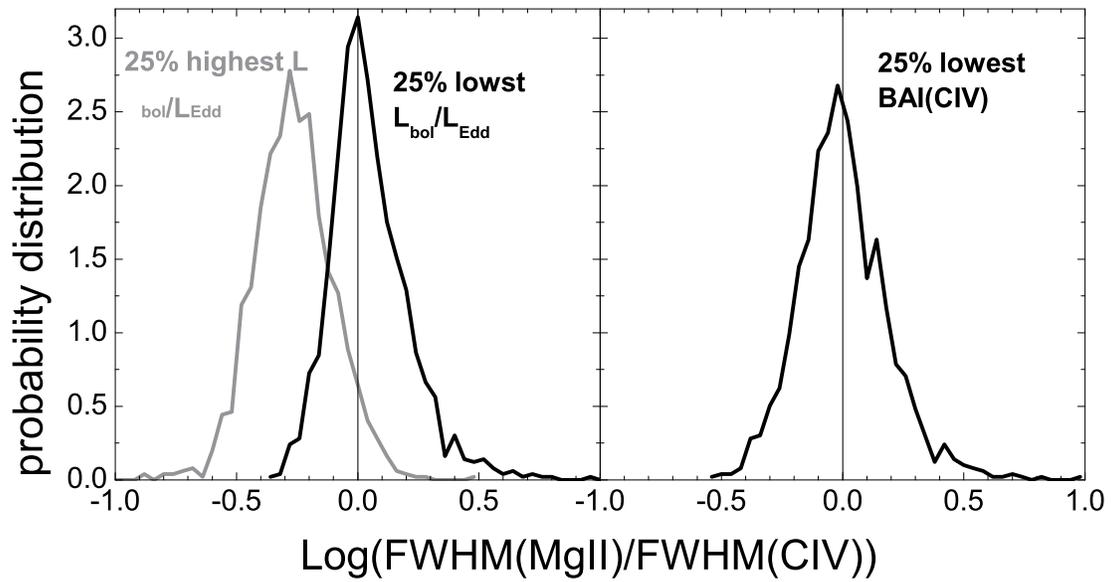}\caption{Left panel: the
probability distributions of $\log{\rm FWHM(\mgii)/FWHM(\civ)}$
for sample A (gray curve) and sample B(black curve). Right panel:
the probability distribution for the 25\% lowest BAI(\civ)
subsample.}\label{fig_fr}
\end{figure}

Therefore, our finding reconciles naturally previous contradictory
results. The \civ\ region tends to be dominated by outflows in
high Eddington ratio AGNs, and dominated by normal gravitationally
bound BELR in low Eddington ratio AGNs. The emission strength from
these two regions varies with \lr\  in opposite directions, and
BAI is a measure of their radio. For low-BAI AGNs, in which the
outflow component is trivial, \civ\ is emitted primarily from the
gravitationally bound region and is expected have a similar line
width to \mgii. This is supported by the probability distribution
of $\log({\rm FWHM(\mgii)/FWHM(\civ)})$ for the 25\% lowest-BAI
AGNs, as shown in the right panel of Fig. \ref{fig_fr}.
Interestingly, we find that the distributions of high- and
low-\lr\ AGNs (sample A and B) on the FWHM-BAI plane join at
BAI$\sim0.6$ (Fig. \ref{fig_fb}). At this value the emission from
the two regions may be comparable, and also the distributions of
BAI for all the AGNs peak (Fig. \ref{fig_udis}). We thus conclude
that the gravitationally bound and radiation driven \civ\ emitting
regions coexist in most of the AGNs (see also Richards et al
2011), and there is no abrupt transition from one type to the
other.

\subsection{Models for Two \civ\ Regions}

Extensive efforts have already been devoted to understanding how
an outflow is launched from an accretion disk (e.g. Arav et al.
1994; Murray et al. 1995; Proga, Stone \& Kallman 2000; Everett
2005). These studies showed that the radiation force arising from
line absorption of the central UV continuum can accelerate
outflows to velocities up to 10000-20000\kms. In particular,
Murray et al. (1995) derived an approximate mass loss rate carried
by outflow (their equation 9), that is equivalent to the strength
of outflow, and found that it increases with the Eddington ratio
(see also Proga et al. 2000). This is consistent with our finding
here in observation. Another important parameter that also has
impact on outflows is the strength of the X-ray emission relative
to the UV (denoted as $\alpha_{\rm ox}$). A hard ionizing
continuum can over-ionize outflows so that the acceleration
efficiency of line absorption is  suppressed; whereas a soft
ionizing continuum would allow the launch of a strong wind (see
e.g. Murray et al. 1995). Recently, Richards et al. (2011) found
that AGNs with larger \civ\ blueshifts are apparently weaker in
X-ray relative to UV (see also Gibson et al. 2008), consistent
with the theoretical expectation. Both the above facts strongly
favor the line driven outflow model.

If $\alpha_{\rm ox}$ decreases with increasing \lr\, the
dependencies of the \civ\ blueshift on \lr\ and $\alpha_{\rm ox}$
are possibly induced by the same underlying causal process. Recent
studies have revealed a complicated relationship between the two
parameters. Kelly et al. (2008) found that $\alpha_{\rm ox}$
decreases with $L_{\rm UV}$/\ledd, but increases with $L_{\rm
X}$/\ledd, where $L_{\rm UV}$ and $L_{\rm X}$ are the UV and X-ray
luminosity, respectively. Such a difference might be ascribed to
the bolometric corrections of these two bands, that change with
\lr\ in different ways (Vasudevan \& Fabian 2007). Vasudevan \&
Fabian (2007) directly calculated \lbol\ by fitting the broad
spectral energy distributions and didn't find any relationship,
while they pointed out that $\alpha_{\rm ox}$ cannot characterize
the full ionizing continuum. Shemmer et al. (2008) also found that
the dependence of $\alpha_{\rm ox}$ on \lr\ is rather weak. They
suggested that this correlation is probably a secondary effect of
the correlations of $L_{\rm UV}$ with \lr\ and $\alpha_{\rm ox}$.
Thus the outflow strength might be affected by least two
`independent' quantities: the Eddington ratio and the relative
strength of the ionizing continuum.

One prediction of the line-driven outflow model is that the
correlation between the outflow strength and \lr\ depends on the
ionizing continuum. In Fig. \ref{fig_uledd}, we show BAI against
\lr\ for the two subsamples that comprise the 25\% largest
EW(\civ) and 25\% smallest EW(\civ) AGNs respectively. One can
find that the correlation between BAI and \lr\ is apparently
stronger in the small EW sample than in the large EW sample. Since
EW(\civ) is strongly correlated with $\alpha_{\rm ox}$ (e.g. Wang
et al. 1998; Wu et al. 2009), it is reasonable to expect that the
small-EW AGNs are relatively X-ray weak compared to the large-EW
ones. The result shown here thus provides another piece of
possible evidence to support the model.

A viable model for the \civ\ emitting region must also account for
the weakening of the normal \civ\ BELs in high-\lr\ AGNs. One
possible mechanism is the change of the ionization state of BEL
clouds, as discussed in Leighly (2004). In that model, the
ionizing continuum is filtered through an outflow before reaching
the normal BELR. The outflow absorbs the photons that can produce
highly ionized ions, such as C$^{+3}$ and He$^{+2}$. As the
outflow becomes stronger, the continuum filtering is more severe
and high-ionization lines from the normal BELR become weaker. A
shortcoming of this model is that it cannot explain the weakness
of the low-ionization line \mgii\ in large BAI (strong outflow) or
high \lr\ AGNs (Fig. \ref{fig_uwe}). Because the ionization state
of the outflow is high,  it can only absorb photons in the helium
continuum, but is transparent to low-energy ionizing photons
(Leighly 2004). Note that a high ionization state is required for
the outflow as there is no significantly blueshifted component in
low-ionization lines. Nevertheless, the ionization state-SED
scenario cannot be ruled out. The low-energy ionizing continuum
may be just intrinsically weak when a strong outflow is launched.

Another possible mechanism is that the amount of BEL clouds varies
with \lr. If radiation pressure can expel BEL clouds out of the
BELR (Dong et al. 2009; 2011, see also Ferland et al. 2009), the
suppression of the normal BEL in high-\lr\ AGNs can be easily
explained. Since BEL clouds are usually dust free, the radiation
pressure arises mainly from three processes: Thomson scattering,
resonance scattering and ionization absorption. Thomson scattering
is not favored since it is only important in AGNs with supper
Eddington luminosities. Resonance scattering is neither able to
drive clouds, because the internal velocity dispersion of a cloud
is generally small and only tiny fraction of photons can be
scattered. Recently, Fabian et al. (2006; see also Marconi et al.
2008) found that the radiation pressure arising from absorption of
ionizing photons cannot be neglected. Assuming the fraction,
$f_{\rm a}$, of ionizing flux is absorbed by a cloud with column
density $N_{\rm H}$, we can derive the ratio of the radiation
force (due to ionization absorption) to the gravitational force on
the cloud:
\begin{equation}
R_f=\frac{F_r}{F_g}=\frac{\frac{L_{\rm ion}f_{\rm a}}{4\pi r^2
c}}{\frac{G \mbh \colnh m_p}{r^2}} =\frac{b f_{\rm a}}{\st
\colnh}\frac{L_{\rm bol}}{L_{\rm Edd}}
\end{equation}
where $G$, $c$ and \st\ are gravitational constant, light speed
and Thomson cross section, $L_{\rm ion}$ the luminosity of
ionizing continuum, $b=L_{\rm ion}/L_{\rm bol}$, and $r$ the
distance from the cloud to the central source. When $R_f>1$,
the radiation force overcomes the gravity of the
central black hole, and the clouds can escape from the BELR. It
sets a lower limit of the column density of clouds that can
survive: $f_{\rm a}N_t\lr$, where $N_t=b/\st$.

There are two parameters, $f_{\rm a}$ and $N_t$, to be determined.
$f_{\rm a}$ is dependent on ionization degree and column density.
For the typical normal BELR, when $N_{\rm H}>
1.2\times10^{21}$cm$^{-2}$, that is the column density at the
hydrogen ionization front, the cloud is optically thick to the
ionizing continuum and $f_{\rm a} \approx 1$ (Ferland 1999;
Marconi et al. 2008). For $N_{\rm H}< 1.2\times10^{21}$cm$^{-2}$,
ionization absorption becomes ineffective, i.e. $f_{\rm a} \approx
0$. We then give a rough estimate of $N_t$ assuming the ionizing
continuum seen by the cloud is the same as what we observe.
Approximating the continuum at wavelength short of 1200\A\ as a
power law with an index $\alpha_{\nu}=-1.57$ (Zheng et al. 1997;
Telfer et al. 2000), $L_{\rm ion}$ is given by integrating over
all ionizing frequencies(i.e. above the frequency of Lyman edge):
$L_{\rm ion}\simeq 1.5\lambda L_{\lambda}(1200\A)$. Then
approximating the UV continuum between 1200\A\ and 3000\A\ as a
power law with an index $\alpha_{\lambda}=-1.54$ (Vanden Berk et
al. 2001), $\lambda L_{\lambda}(1200\A) \simeq 1.64 \lambda
L_{\lambda}(3000\A)$. Adopting $L_{\rm bol}\simeq 5.9 \lambda
L_{\lambda}(3000\A)$ (McLure \& Dunlop 2004), we obtain $b\approx
0.41$ and $N_t\approx 6\times10^{23}$cm$^{-2}$.

As long as $\lr \gs 1.2\times10^{21}/6\times10^{23}=
2\times10^{-3}$,  clouds with column density in the range between
$1.2\times10^{21}$cm$^{-2}$ and $N_t\lr$ can be blown away from
the BELR. That is to say, the amount of BEL clouds that can
survive near the central engine drops with increasing \lr.
Increasing \lr\ up to 0.1, the upper limit of the column density
of expelled clouds increases to $6\times 10^{22}$cm$^{-2}$; where
clouds can effectively emit the low-ionization line \mgii. Note
that the estimation of $N_t$ is inexact, since it depends on
whether the ionizing emission is isotropic, or is blocked by other
BEL clouds (Gaskell 2009) and/or other structures (such as
outflows, Leighly 2004). Nevertheless, the mechanism proposed here
gives a promising explanation to the suppression of both the \civ\
and \mgii\ emission from the normal BELR in high-\lr\ AGNs.

\subsection{Connection Between \civ\ Blueshift and BAL}\label{sec_cbo}

One of the most prominent spectral features imprinted by outflows
is the broad absorption line. We refer to outflows associated with
BAL as BAL outflow, while that associated with blueshifted BEL as
BEL outflow. The comparison below suggests that BAL and BEL
outflows may represent the same physical component.

Richards et al. (2002) found that the emission line features, such
as the \ciii\ line complex\footnote{\ciii\ line complex is composed of
\ciii, \siiii\ and \aliii.}, \heii\ and the red wing of \civ, are
very similar between the composites of BAL AGNs and non-BAL AGNs
with large \civ\ blueshift. The broad band SED also displays
a similar trend. For a given UV luminosity, non-BAL AGNs with large
blueshifts tend to be X-ray weak compared to those with small
blueshifts (Richards et al. 2011). The X-ray emission of BAL AGNs
is usually severely absorbed, and thus cannot be directly used for a
comparison. After correction for absorption, Fan et al.
(2009) found that the intrinsic X-ray emission of BAL AGNs is on
average weaker than that of non-BAL AGNs of the same UV
luminosity. In particular, they found that the intrinsic X-ray strength
is anti-correlated with the absorption strength of the \civ\ BAL for
BAL AGNs. It is consistent with the correlation for the \civ\ BEL.

More similarities can be found when looking at the correlations of
the outflow properties with \lr. In this work, we find that the
BEL outflow is stronger at a higher \lr\ (Fig. \ref{fig_uledd}).
Similar correlations between the BAL properties and \lr\ have been
reported previously. Boroson (2002) has shown that BAL AGNs tend
to occupy the extreme end of the Boroson \& Green (1992)
Eigenvector 1, that is thought to be driven by the Eddington
ratio. Recently, Ganguly et al. (2007) found the fraction of BAL
AGNs, an indicator of the average covering factor of outflow,
increases with the Eddington ratio(see also Zhang et al. 2010).
They also found there appears to exist an overall upper-envelope
of increasing $v_{\rm max}$ of BALs with increasing Eddington
ratio. The correlations for the BAL outflows are consistent with
but weaker than those for the BEL outflows. There may be a simple
reason for this: BAL troughs hold only the information of outflow
along the line of sight, and thus may be sensitive to the local
structure; while BEL is an integral of emission over entire volume
of the outflow and represents the overall properties.

The similarity between the BEL and BAL outflows can also be found
directly from their own properties. The first  is the
maximal velocity, $v_{\rm max}$. Gibson et al. (2009) have obtained
the distribution of $v_{\rm max}$ for BAL AGNs, which
ranges from $<$5000\kms\ to $>$20000\kms, with a mean value of
$\sim$12000\kms. The maximal velocity of a BEL outflow for individual AGN
is hard to estimate. We estimate the average $v_{\rm max}$ of the
BEL outflow from the difference spectra of the largest-BAI and
smallest-BAI composites (Fig. \ref{fig_ex}). It is about
11000\kms, similar to the BAL outflow.
We then compare their ionization states. As discussed above, the absence of
blueshifted \mgii\ suggests that the BEL outflows contain very
few Mg$^+$ ions. The same condition also appears in the BAL outflows.
About only 10\% of the BAL AGNs exhibit \mgii\ BAL (e.g Trump et al. 2006;
Zhang et al. 2010), suggesting the absence of Mg$^+$ ions in
most of the BAL outflows.

Given the  similarities shown above,  the BAL outflows may also be
responsible for producing the observed blueshifted BEL, at least
for the strongest blueshifted BEL. When the line of sight to the
continuum source intersects the outflow, BALs are produced in the
AGN spectrum; otherwise, the AGN appears as a non-BAL AGN with
blueshifted \civ. The frequency of occurrence of  BAL in AGN
spectra, the strength of BAL trough, and the blueshift of \civ\
are all enhanced as the outflow is boosted with increasing \lr\
and/or  decreasing the intrinsic X-ray emission. A viable outflow
model must simultaneously account for both, the absorption and
emission of \civ. Future studies on AGN outflows can be proceed
from these two different approaches. For example, it is possible
to use reverberation mapping results of the blueshifted \civ\ line
to estimate the distance of the BAL outflow to the central source,
an important parameter for understanding the launch of the outflow
(the difficulty lies in how one separates variation of the outflow
component from that of the normal BEL). One may also use the X-ray
emission of AGNs with strongly blueshifted \civ, rather than the
entire sample, to characterize the intrinsic X-ray emission of BAL
AGNs; the latter is generally strongly absorbed and hard to study
directly. In addition, we note that resonance scattering of the
continuum and emission lines by ions, such as C$^{3+}$ and
N$^{4+}$, in outflows may also contribute to the observed emission
(e.g. Wang et al. 2010). That is to say, for proper modeling of
the emission lines, especially \civ\ and \nv\ with large
blueshifts (a strong outflow and a weak normal BELR), the
scattering emission must be taken into consideration.

\section{Summary}\label{sec_sum}

There are two kinds of models proposed  to describe the kinematics
and structure of the BELR of high-ionization lines (e.g. \civ) in
AGNs, namely, the outflow and gravitationally bound BELR,
that are apparently mutually contradictory. In this paper we attempt to uncover the
underlying physical process responsible for this difference, using
4963 AGNs in the redshift range $1.7<z<2.2$ selected from SDSS
DR5.

We introduce a blueshift and asymmetry index (BAI) to measure the
deviation of the \civ\ line from an unshifted and symmetric
profile. BAI is defined as the flux ratio of the blue part to the
total profile, where the blue part is the part of the \civ\ line
at wavelengths short of 1549.06\A, the laboratory rest-frame
wavelength of the \civ\ doublets. BAI actually measures the
combined effect of asymmetry and shift of a line profile, which is
generally treated separately in the literature. Since both
asymmetry and line shift may be caused by the same physical
process, BAI is the best choice for our purpose.

We confirm previous findings that the \civ\ BEL is generally
blueshifted with respect to the systemic redshift, and there exist
significant correlations among the BAI, the line width (FWHM) and
the rest equivalent width (EW) of \civ. For comparison, we
investigate the same relationships among the BAI of \civ, the FWHM
and EW of the low-ionization line \mgii. Dramatic differences are
found between these two lines. For the \mgii\ line, FWHM is
positively correlated with EW, and inversely correlated with BAI.
While for \civ, FWHM is anti-correlated with EW, and the
correlation between FWHM and BAI is complex and significantly
different from that for \mgii. Given that line profile reflects
the kinematic properties of the emitting gas, our results
demonstrate clearly that, in general, the \civ\ line comes from a
region with the structure and kinematics fundamentally different
from those of \mgii, which is thought to be gravitationally bound.

We find a prominent correlation between BAI and the Eddington
ratio, that is consistent with theoretical expectation. This
correlation, together with the previously known correlation
between \civ\ blueshift and the X-ray to UV flux ratio, advocates
strongly the view that outflows are driven by resonance line
absorption. Interestingly, there exist a number of similarities
between blueshifted emission lines and broad absorption lines in
AGNs in, for instance, the maximal velocity, ionization state and
the correlations with the Eddington ratio and with the X-ray to UV
flux ratio. These suggest that the same outflow produces these two
different phenomena observed.

We select two subsamples, one composed of the 25\% highest and one
of the 25\% lowest Eddington ratio AGNs. In the high Eddington
ratio subsample, FWHM(\civ) is positively correlated with BAI and
anti-correlated with EW(\civ). Both correlations are opposite to
those for \mgii\ and can be understood in terms of outflow.
Whereas, for the low Eddington ratio subsample, these two
correlations for \civ\ become marginal or absent. \civ\ overlaps
almost completely  with \mgii\ in the parameter space. In
particular, the line width of \civ\ is, on average, the same as
that of \mgii. We thus conclude that the \civ\ line in this
subsample is emitted by optically-thick gas driven by gravity,
similar to the \mgii\ line.

Our results further suggest that the two \civ\ emitting regions,
gravitationally bound and radiation driven, coexist in most of the
AGNs. The emission strengths of these two regions vary smoothly
with \lr\ in opposite directions. The \civ\ emission is generally
dominated by outflows at high Eddington ratios, while it is
primarily emitted from the normal gravitationally bound BELRs at
low Eddington ratios. This explanation naturally reconciles the
contradictory views proposed in previous studies. Viable models
are also discussed that can account for both, the enhancement of
outflow emission and the suppression of the normal BEL, in AGNs
with high Eddington ratios.

\acknowledgements We thank the referee, Gordon Richards, for his
helpful suggestions that improved the paper. This work is
supported by NSFC (11073017, 11033007, 10973013, 10973012,
11073019), 973 program (2007CB815405, 2009CB824800) and the
Fundamental Research Funds for the Central Universities. Funding
for the SDSS and SDSS-II has been provided by the Alfred P. Sloan
Foundation, the Participating Institutions, the National Science
Foundation, the U.S. Department of Energy, the National
Aeronautics and Space Administration, the Japanese Monbukagakusho,
the Max Planck Society, and the Higher Education Funding Council
for England. The SDSS Web site is http://www.sdss.org/.

\appendix{}

\section{Notes on the asymmetry and shift of line profile}

The asymmetry and shift of line profile
are generally treated separately in the literature (e.g. Baskin \& Laor 2005).
Our blueshift and asymmetry index (BAI) is actually
a  combination of these two properties. In this Appendix,
we present the line-shift index (SI) and asymmetry index (AI) of the
\civ\ line profile and their comparisons with BAI.
Both parameters are measured from the composite profiles
of our two-Gaussian model.
SI is defined as the shift of
the line peak with respect to that in the AGN rest frame.
Positive values of SI indicate blueshift.
While AI is the flux ratio of the blue part to the total profile,
where the blue part is the part of line profile
at wavelengths short of the \emph{peak wavelength}.
The measurement of AI is therefore independent of the redshift errors of AGNs.

\begin{figure}
\epsscale{1}\plotone{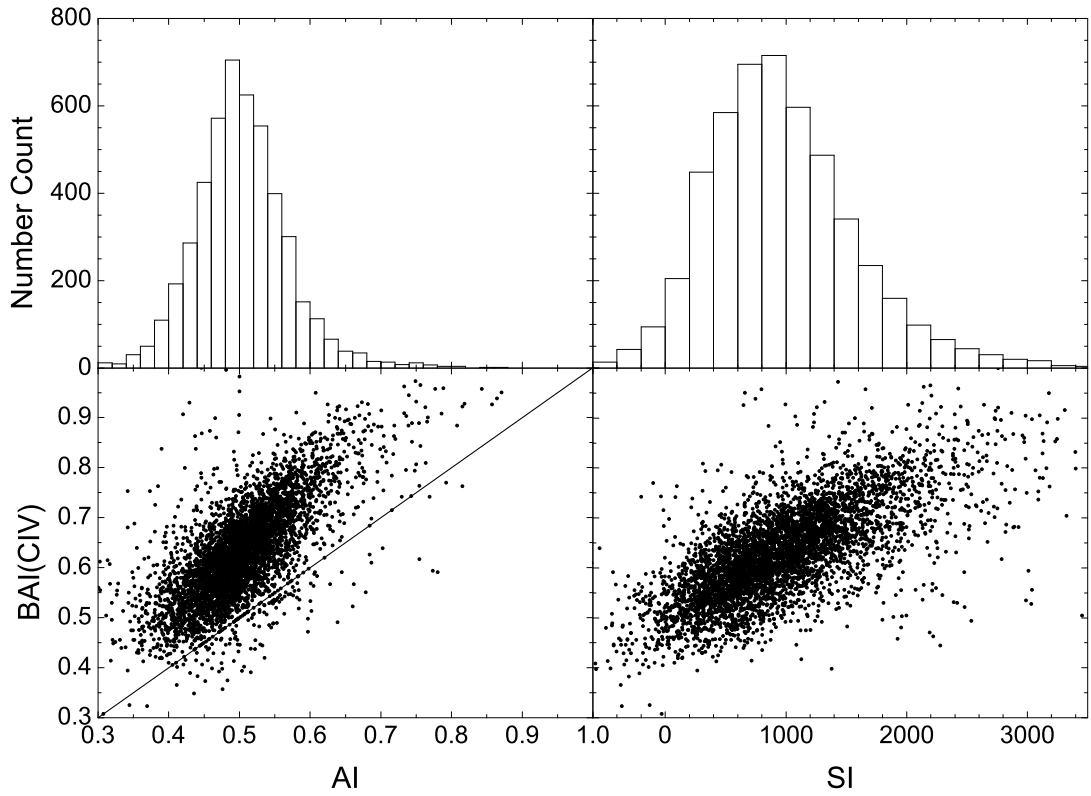}\caption{Upper-left panel: the distribution of asymmetry index (AI). Upper-right
panel: the distribution of line-shift parameter (SI). Lower-left panel: BAI versus AI. Lower-right panel: BAI versus SI.}\label{fig_asy}
\end{figure}

In the upper panels of Fig. \ref{fig_asy}, we show the
distributions of AI and SI. The median of SI is about 898\kms, in
agreement with Richards et al. (2011). The median of AI is about
0.50, suggesting that there is no strong preference for red or
blue asymmetries. This is consistent with the result in Marziani
et al. (1996, their Figure 4, see also Sulentic et al. 2000a), but
different from that in Baskin \& Laor (2005), who found that the
\civ\ lines tend to be blue asymmetric. To derive the line
parameter, Baskin \& Laor fitted the local continuum with a
power-law in two wavelength windows near 1470\A\ and 1620\A.
However, it is well known that there is a prominent bump near
1620\A\ (see e.g. Nagao et al. 2006). Adopting 1620\A\ as a
continuum window can lead to systematic overestimation of AI.

AI and SI are strongly correlated with BAI (the lower panels of
Fig. \ref{fig_asy}). The Spearman rank coefficients for these two
correlations are both about $r_s=0.7$, suggesting that redshift
errors do not cause significant bias in BAI, at least for our
sample with high S/N data. We use AI or SI, instead of BAI, to
repeat our analysis. Interestingly, all the correlations shown in
this paper retain, albeit becoming somewhat weaker. This means
that we can draw the same conclusions by using any of these three
parameters. This further suggests that the line asymmetry and line
shift are caused by the same process, as is discussed in the main
text. Since BAI is a combination of AI and SI, BAI is the best
choice for our purpose here.

Moreover, we also attempt to use the often-used asymmetry index,
such as the shift between the centroids at $1/4$ and $3/4$ maximum in units of
FWHM (see Sulentic et al. 2000a; Baskin \& Lasor 2005).
Very similar results are obtained.

\end{document}